**Enormous enhancement of p-orbital magnetism and band gap in the lightly doped carbyne**

C.H.Wong[*,1], R. Lortz[2], A.F.Zatsepin[1]

[1]Institute of Physics and Technology, Ural Federal University, Yekaterinburg, Russia.

[2]Department of Physics, The Hong Kong University of Science & technology, Clear Water Bay, Kowloon, Hong Kong

[*]ch.kh.vong@urfu.ru

This paper presents a path to tailor adapted magnetic and optical properties in carbyne. Although p-orbital magnetism is generally much weaker than d-orbital magnetism, we demonstrate that the charge fluctuation of the free radical electrons triggered by a time-varying electric dipole moment leads to enormous p-orbital magnetism. By introducing 25% arsenic and 12.5% fluorine into the monoatomic carbon chain, the magnetic moment of the arsenic atom reaches $2.9\mu_B$, which is ~1.3 times stronger than magnetic moment of bulk Fe. This magnetically optimized carbyne composite carries an exchange-correlation energy of 22meV (~270K). On the other hand, we convert the carbyne (in beta-form) from metallic to a semiconducting state by using anionic dopants. After doping 12.5% nitrogen and 12.5% oxygen into the beta-carbyne, the semiconducting gap of this composite is optimized at 1.6eV, which is 1.4 times larger than the band gap of bulk silicon.

An electron carries a magnetic dipole moment as an intrinsic spin property. In addition, when the electron moves around a circular orbit in an atom, it generates an orbital magnetic moment.. Materials with strong magnetic moments are mainly found in heavy atoms with partially filled d or f shells in which all magnetic quantum numbers add up to a large magnetic moment. Although carbon materials do not contain electrons in d-shells, the circular motion of the electrons can activate p-orbital magnetism when the p-orbits effectively overlap [1,2,3]. The disappearance of orthogonal interaction in a one-dimensional system allows the resulting orbital overlap of p-shells concentrating on the chain axis, so that the p-orbital magnetism is more favorable in a one dimensional structure [1,4,5]. An early investigation on the monoatomic lithium or sodium nanowires showed that the strength of p-orbital ferromagnetism depends on the bond length and the bond angle [1,4]. When the p-shells begin to overlap, a tiny increment of the orbital area causes the p-orbital magnetism [1]. A. Bergara *et al* have verified that the optimized magnetic moment of the p-shell corresponds to a maximum overlap between the p-shells [1]. If the orbital-overlap is too strong, the circulation motion of the electron is disturbed by a strong electrostatic repulsion which destroys the magnetic moment of p-shell [1,5].

Although the magnetic moment of p-shell may be weak [6], the electrons that move less freely on p-shells allow a strong exchange coupling [5]. An experimental observation confirmed that the p-orbital ferromagnetism of the undoped monoatomic carbon chain persisted up to 400K, even though the magnetic moment of carbon is only ~0.2$\mu_B$ [5]. While a large exchange interaction is possible in p-orbital magnetism [5,7], this will be a promising tool to design the next generation of spintronic devices when the magnetic moment of p-shells is significantly enhanced. Dopant play an important role in p-orbital magnetism [8]. The use of anionic dopants increases the local magnetic moment of p-shell in the carbon chain to ~1.4 $\mu_B$ due to charge fluctuations of the free radial electron [5]. Nevertheless, all reported p-orbital magnetisms of carbon material are much weaker than the d-orbital magnetism of transition metals unfortunately [1,5,8].

On the other hand, both Monte Carlo simulation and first principles calculations predict that the monoatomic carbon chain remains in the beta phase (⋯=C=C=C=⋯) and behaves as metal at room temperature [9,10]. This is the reason, beta-carbyne cannot attract much attention from the semiconductor industries. In order to solve this problem, we have set ourselves the goal of modifying the carbyne to that it is semiconducting at room temperature. In addition, we will increase the magnetic moment of the p-shell until it exceeds the magnetic moment of bulk iron.

The initial bond distance between the adjacent carbon atoms along the chain axis is 1.28pm, the pivot angle 180 degrees. The monoatomic carbon chains are separated laterally by 1nm. After geometric optimization, the optical band gap and the magnetic moment are calculated by the spin-unrestricted GGA-PW91 functional [11,12]. The chemical formula of the carbyne composite is abbreviated as 'A(12.5%) + B(12.5%) + carbyne(75%)', where A or B refer to dopants (unless otherwise stated), with the schematic diagram shown in Fig.1. The dopants belonging to Group V, VI & VII are abbreviated as V-dopants, VI-dopants and VII-dopants, respectively. The dopant produces a kink as a local curvature and the asymmetric bond distances are compared by $R_A = \frac{L_{C2C3}}{L_{C1C2}}$ or $R_B = \frac{L_{C5C6}}{L_{C4C5}}$ where $L_{C2C3}$, $L_{C1C2}$, $L_{C5C6}$, $L_{C4C5}$ are the bond lengths of the carbon atom correspondingly.

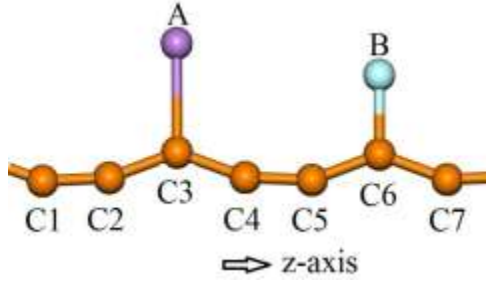

FIG.1. The local region of "A(12.5%) + B(12.5%) + carbyne(75%)" along the chain axis. The dopants are noted as A and B. The repeated unit contains seven carbon atoms, each labelled as C1, C2, C3, C4, C5, C6, C7.

Although beta-carbyne is predicted to be metallic below 500K [9,10], doping anionic atoms into beta-carbyne can successfully open the semiconducting gap. Compared to the V-doped carbyne and VII-doped carbyne, the VI-doped carbyne has a more notable effect in opening an energy gap, as listed in Table 1. Therefore, we focus on tuning the optical properties of the VI-doped carbyne using different atomic radii. However, Table 2 shows that the semiconducting gap is narrower when using a larger VI-dopant. The doping light elements in the beta-carbyne look promising to increase the band gap. Despite the band gap of "N(12.5%) + N(12.5%) + carbyne (75%)" reaches 1.1eV only, Fig.2a demonstrates that the band gap of "N(12.5%) + O(12.5%) + carbyne (75%)" can be tremendously optimized to 1.6eV. In contrast, Table 3 illustrates that the band gap of "N(12.5%) + F(12.5%) + carbyne (75%)" is decreased to 0.7eV when fluorine is doped. The O-doped, S-doped, Se-doped or F-doped carbyne composite show no magnetism at all. We observe that the band gap is usually inversely proportional to $R_A$ & $R_B$.

| Sample: A + B + carbyne | $E_g$(eV) | M($\mu_B$) | M($\mu_B$) | $R_A$(Å) | $R_B$(Å) |
|---|---|---|---|---|---|
| N(12.5%) + N(12.5%) + carbyne (75%) | 1.076 | N: 0.921 | N: 0.921 | 1.156 | 1.156 |
| O(12.5%) + O(12.5%) + carbyne (75%) | 1.230 | O: 0 | O: 0 | 1.178 | 1.178 |
| F(12.5%) + F(12.5%) + carbyne (75%) | 0 | F: 0 | F: 0 | 1.090 | 1.090 |

Table 1: The optical and magnetic properties of the homogenously doped carbyne composites. $E_g$ is the band gap, M is the magnetic moment.

| Sample: X + Y + carbyne | $E_g$(eV) | M($\mu_B$) | M($\mu_B$) | $R_A$ | $R_B$ |
|---|---|---|---|---|---|
| O(12.5%) + O(12.5%) + carbyne (75%) | 1.230 | O: 0 | O: 0 | 1.178 | 1.178 |
| O(12.5%) + S(12.5%) + carbyne (75%) | 0.500 | O: 0 | S: 0 | 1.162 | 1.140 |
| O(12.5%) + Se(12.5%) + carbyne (75%) | 0.341 | O: 0 | Se: 0 | 1.157 | 1.129 |

Table 2: The band gaps and magnetism in various VI-doped carbyne composites.

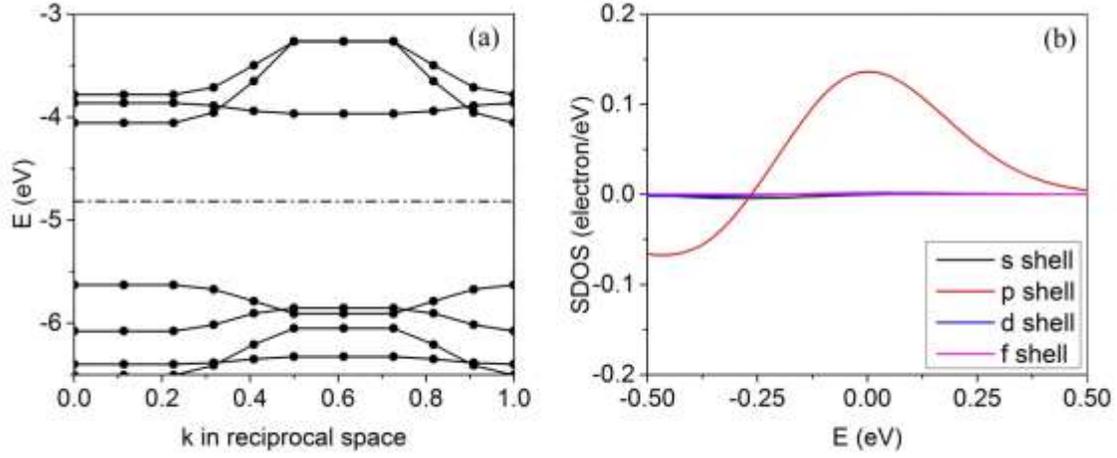

FIG.2. The optical and magnetic properties of "N(12.5%) + O(12.5%) + carbyne(75%)". (a) band structure. The dash line shows the actual Fermi level. (**b**) the differential spin density of states per atom. The Fermi level is shifted to 0eV for convenience.

Although the use of V-dopant and VII-dopant is not an effective approach to increase the band gap of beta-carbyne, the dopants in these two groups give great hope for the activation of a giant p-orbital magnetism in which the magnetic moments are listed in Table 3. All calculated magnetic moments come from p-orbital magnetism and we show the magnetism of "N(12.5%) + O(12.5%) + carbyne(75%)" in different shells as an example in Fig.2b.

| Sample: X + Y + carbyne | $E_g$(eV) | M($\mu_B$) | M($\mu_B$) | $R_A$ | $R_B$ |
|---|---|---|---|---|---|
| N(12.5%) + N(12.5%) + carbyne (75%) | 1.076 | N: 0.921 | N: 0.921 | 1.156 | 1.156 |
| N(12.5%) + O(12.5%) + carbyne (75%) | 1.574 | N: 0.909 | O: 0 | 1.167 | 1.172 |
| N(12.5%) + F(12.5%) + carbyne (75%) | 0.722 | N: 1.371 | F: 0.036 | 1.132 | 1.099 |

Table 3: The optical and magnetic properties of the carbyne composites

Table 4 shows the p-orbital magnetism of the V & VII doped carbyne composites. A stronger magnetic moment is observed in a heavier atom. However, the doping of a heavier VII-dopant surprisingly weakens the magnetic moment of the V-dopant. The arsenic atom in the "As(12.5%) + F(12.5%) + carbyne (75%)" holds a magnetic moment as strong as 1.8$\mu_B$, which almost corresponds to the bulk Fe's moment of 2.2$\mu_B$. When two arsenic atoms are bonded to the same carbon atom to form "As(25%) + F(12.5%) + carbyne (62.5%)", the magnetic moments of arsenic are boosted to 2.86μ$_B$ and -1.75μ$_B$ respectively. The exchange correlation energy of this magnetically optimized carbyne composite is 22meV (~270K). Fig.3a shows that the band gap of "As(25%) + F(12.5%) + carbyne (62.5%)" is only ~0.3eV. Fig.3b confirms that its giant magnetism is again due to p-shells. All magnetic moments of carbon in our materials under investigation are ~0.2μ$_B$. All our attempts to increase the band gap and magnetic moment beyond our predicted 1.6eV and 2.9μ$_B$ have failed so far.

| Sample: X + Y + carbyne | E$_g$(eV) | M(μ$_B$) | M(μ$_B$) | R$_A$ | R$_B$ |
|---|---|---|---|---|---|
| N(12.5%) + F(12.5%) + carbyne (75%) | 0.722 | N: 1.371 | F: 0.036 | 1.132 | 1.099 |
| N(12.5%) + Cl(12.5%) + carbyne (75%) | 0.670 | N: 1.352 | Cl: 0.060 | 1.129 | 1.097 |
| N(12.5%) + Br(12.5%) + carbyne (75%) | 0.647 | N: 1.345 | Br: 0.072 | 1.129 | 1.094 |
| | | | | | |
| P(12.5%) + F(12.5%) + carbyne (75%) | 0.339 | P: 1.741 | F: 0.022 | 1.094 | 1.086 |
| P(12.5%) + Cl(12.5%) + carbyne (75%) | 0.320 | P: 1.690 | Cl: 0.051 | 1.091 | 1.078 |
| P(12.5%) + Br(12.5%) + carbyne (75%) | 0.311 | P: 1.666 | Br: 0.076 | 1.091 | 1.074 |
| | | | | | |
| As(12.5%) + F(12.5%) + carbyne (75%) | 0.266 | As: 1.803 | F: 0.016 | 1.087 | 1.083 |
| As(12.5%) + Cl(12.5%) + carbyne (75%) | 0.251 | As: 1.726 | Cl: 0.048 | 1.072 | 1.082 |
| As(12.5%) + Br(12.5%) + carbyne (75%) | Unstable structure | | | | |

Table 4: The energy gaps and magnetic moments of the heterogeneously doped carbyne composites.

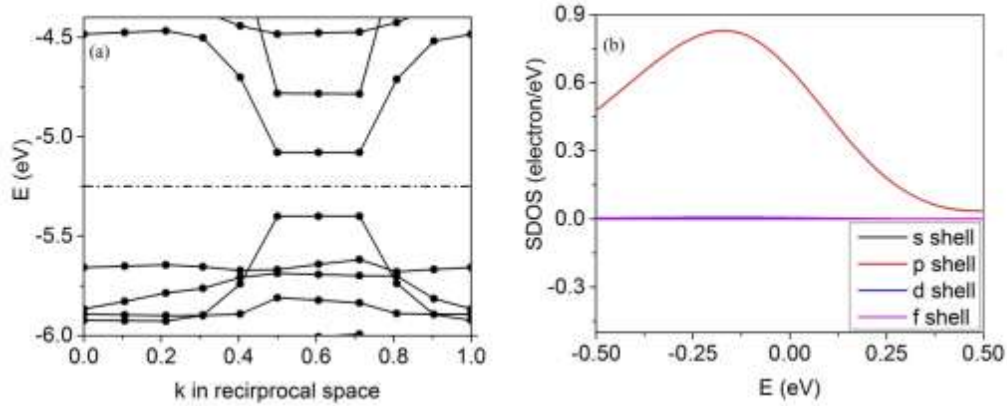

FIG.3. The optical and magnetic properties of "As(25%) + F(12.5%) + carbyne(62.5%). (**a**) band gap is ~0.3eV where the dash line shows the actual Fermi level. (**b**) The differential spin density of states per atom. The Fermi level is adjusted to 0eV for a clearer presentation.

The asymmetric bond lengths associated with the optical phonon [6] are generated along the carbon chains when the $R_A$ or $R_B$ is greater than 1. The "O(12.5%) + O(12.5%) + carbyne (75%)" holds a larger band gap than the "N(12.5%) + N(12.5%) + carbyne (75%)" because the $R_A$ and $R_B$ of the O-doped carbyne are relatively higher than those listed in Table 1. The F-doped carbyne cannot open a band gap because the asymmetric bond length is less pronounced.

Neither LDA nor GGA can provide an accurate band structure of semiconductors [13,14]. The theoretical band gaps computed by the LDA and GGA like PW91, PBE or BYLP are always underestimated [13-17]. In principle, the error of bandgap can be minimized by considering a more complex interaction between electrons. The use of the GW corrected GGA or other hybrid functional is a common approach to bring the band gap closer to the experimental observations [18]. However, there is no unique way to calibrate the DFT functional. The Hamiltonian, boundary conditions and simulation parameters of the DFT calculation depend on the sample [13-18]. Since the experimental band gap of our doped carbyne is still unknown, we consider the GGA-computed band gap to be an approximate predicted value. Although our theoretical band gap is underestimated, our optically optimized sample still shows the band gap at 1.6eV, which may allow a wide variety of optical applications.

Based on the earlier study of the monoatomic Li or Na chain, p-orbital magnetism occurs when the p-shells begin to overlap [1]. Once the orbital overlap exceeds the optimal point, electron repulsion weakens p-orbital magnetism [1]. However, the magnetic moment of the p-shell by modification of the orbital area is only ~0.1$\mu_B$, which is not sufficient to explain why the magnetic moment of our optimized sample is stronger than the d-orbital magnetism in some transition metals [19].

We propose a free-radical model (FR model) to take into account the unusually large p-orbital magnetism [5]. *Our FR model states that the free radical electron dynamics and the quantum degeneracy can be used to amplify p-orbital magnetism. The origin of the magnetic moment is the magnetic field generated by an electron moving around a circular orbit. When the free radical electron moves across the dopant, the free radical electron generates a microscopic current. While the charge fluctuation is present everywhere, the additional B-field is captured by the orbital area of dopant, which is believed to amplify the magnetic moment of the p-shell.*

Fig.4 shows the quantum degeneracies of the O-doped, N-doped and F-doped carbyne composites. Since the electrostatic repulsion between the free radical electrons is strong in the VI-doped carbyne, the free radial electrons move more restrictedly [5]. As a result, p-orbital magnetism is missing in the VI-doped carbyne composites corresponding to our FR model. The doping of N into carbyne is a promising way to activate p-orbital magnetism, as there are vacancies to host the free radical electrons [5]. On the other hand, the free radical electrons in the N-doped carbyne receive an additional kinetic energy from the induced electric dipole [5], as marked in Fig.4c&4d. According to our FR model, the faster the free radical electron drifts, the stronger the microscopic current it induces nearby the V-dopant [5,6]. Despite the two degenerate states in the F-doped carbyne allow the free radical electrons to swap positions in Fig.4e&4f, the velocity of free radical electron is not fast enough to trigger p-orbital magnetism. If the fluorine concentration is reduced by half, a weak magnetic moment is generated at the doped sites because the electrostatic repulsion is minimized [5]. The band gap of "O(12.5%) + O(12.5%) + carbyne (75%)" is the highest among the others in Table 2, since the asymmetric bond length triggered by the optical phonon is more distinct [6].

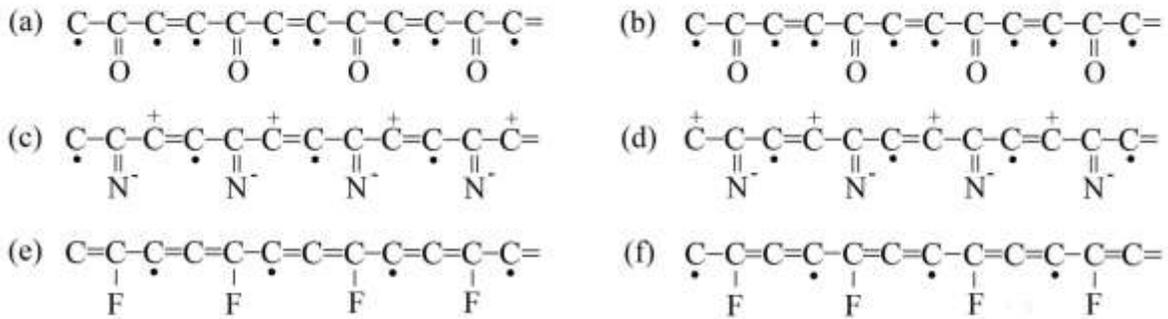

FIG.4. The local regions of the homogeneously doped carbyne. The bond angles are drawn arbitrarily. The '+' and '-' refer to the polarity of electrostatic potential. (a) and (b) are the degenerate states of the O-doped carbyne. (c) and (d) are the degenerate states of the N-doped carbyne. (e) and (f) are the degenerate states of the F-doped carbyne.

According to Table 3 and Table 4, the "N(12.5%) + F(12.5%) + carbyne (75%)" show a stronger p-orbital magnetism than the "N(12.5%) + N(12.5%) + carbyne (75%)". The reason for this is that the free radical electrons '1' and '2' in Figu.5a&5b are dragged to the same N-site, which further amplifies the microscopic current. The p-orbital magnetism is relatively weak in "N(12.5%) + O(12.5%) + carbyne (75%)" due to the large electrostatic repulsion between the free radical electrons (see Fig.5c&5d). Since the kink structure is most pronounced in the "N(12.5%) + O(12.5%) + carbyne (75%)", its semiconducting gap is the largest.

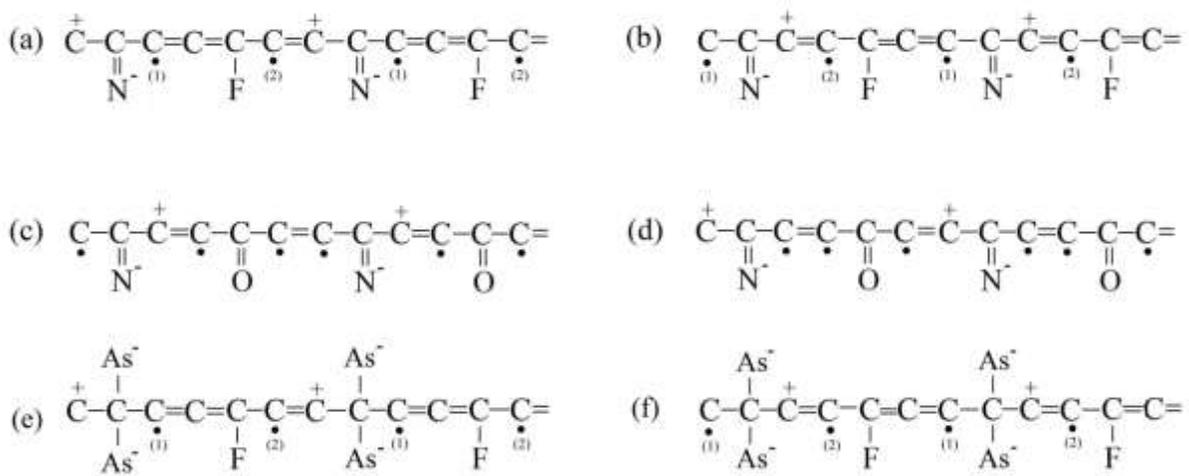

FIG.5. The local regions of the heterogeneously doped carbyne. The bond angles are drawn arbitrarily. The '+' and '-' refer to the sign convention in electrostatic potential. (a) and (b) are the degenerate states of the "N(12.5%) + F(12.5%) + carbyne (75%)". (c) and (d) are the degenerate states of the "N(12.5%) + O(12.5%) + carbyne (75%). (e) and (f) are the degenerate states of the "As(25%) + O(12.5%) + carbyne (62.5%).

After the introduction of a heavier VII-dopant, the magnetic moment of the V-dopant is weaker, as shown in Table 4. While a smaller VII-dopant strengthens the local electrostatic repulsion in the form of a larger kink angle, the enormous electrostatic repulsion from the smaller VII-dopant drives the free radical electron '2' more strongly across the $C^+$ - $N^-$ dipole [5]. In this case, the p-orbital magnetism at the V-dopant is generated more effectively according to our FR model. The magnetic moment of the heavier atom is larger in Table 4 because the orbital area is proportional to the strength of magnetic moment [6].

To investigate whether the electric dipole amplifies the p-orbital magnetism of carbyne, we double the total number of electric dipoles (per repeated unit) in Fig.5e&5f. The strongest magnetic moment of $2.86\mu_B$ occurs in one of the arsenic atoms in the "As(25%) + F(12.5%) + carbyne (62.5%)" because there are two $C^+$ - $As^-$ electric dipoles dragging the free radical electron '1' & '2' simultaneously. As a consequence, the almost doubled microscopic current triggers the unusually large magnetic moment of the arsenic atom. The magnetic moments of these two arsenic atoms point in opposite direction, so that a spin-flip process is expected when the spin-polarized electron is injected across these two arsenic atoms.

We reported a detailed study of the magnetic and optical properties in the lightly doped carbyne composites. The optimization of bond angle, kink angle and charge fluctuation in heterogeneously doped carbyne allows the magnetic moment of the p-shells to exceed the magnetic moment of bulk Fe. Our magnetically optimized sample holds a strong exchange coupling of 22meV (~270K). While the electrical properties of the heterogeneously doped carbon chain are modified from metallic and semiconducting, the optically optimized band gap of 1.6eV may open a potential application in the semiconductor industry.